# Space Development and Space Science Together, an Historic Opportunity

Philip T. Metzger

Florida Space Institute, University of Central Florida,

12354 Research Parkway, Orlando, FL 32826, USA

philip.metzger@ucf.edu

**Abstract**

The national space programs have an historic opportunity to help solve the global-scale economic and environmental problems of Earth while becoming more effective at science through the use of space resources. Space programs will be more cost-effective when they work to establish a supply chain in space, mining and manufacturing then replicating the assets of the supply chain so it grows to larger capacity. This has become achievable because of advances in robotics and artificial intelligence. It is roughly estimated that developing a lunar outpost that relies upon and also develops the supply chain will cost about 1/3 or less of the existing annual budgets of the national space programs. It will require a sustained commitment of several decades to complete, during which time science and exploration become increasingly effective. At the end, this space industry will capable of addressing global-scale challenges including limited resources, clean energy, economic development, and preservation of the environment. Other potential solutions, including nuclear fusion and terrestrial renewable energy sources, do not address the root problem of our limited globe and there are real questions whether they will be inadequate or too late. While industry in space likewise cannot provide perfect assurance, it is uniquely able to solve the root problem, and it gives us an important chance that we should grasp. What makes this such an historic opportunity is that the space-based solution is obtainable as a side-benefit of doing space science and exploration within their existing budgets. Thinking pragmatically, it may take some time for policymakers to agree that setting up a complete supply chain is an achievable goal, so this paper describes a strategy of incremental progress. The most crucial part of this strategy is establishing a water economy by mining on the Moon and asteroids to manufacture rocket propellant. Technologies that support a water economy will play an important role leading toward space development.

1. **Introduction**

Because of recent technological advances it has now become practical and affordable to establish a complete, robotic, industrial supply chain in space, enabling great science while promising tremendous benefits back on Earth [1]. Some of the benefits (mostly in space) occur in the early phase of establishing this industry, while more dramatic benefits occur after it becomes self-sufficient so that no further material need be launched from Earth and it can be scaled-up to great throughput. In this paper the end-state shall be called a Self-sufficient Replicating Space Industry, or SRSI. The main challenge for this concept is neither technology nor cost but simply convincing people it is realistic. In the 1970s Gerard K. O'Neill proposed orbiting space colonies, each with 10,000 residents who would manufacture solar power stations to beam clean energy to Earth at a profit. Senator William Proxmire said of the concept, "It's the best argument yet for chopping NASA's funding to the bone. As Chairman of the Senate Subcommittee responsible for NASA's appropriations, I say not a penny for this nutty fantasy..." [2]. Likely, many people will have the same reaction to a program of bootstrapping SRSI.

To be pragmatic, we may consider this to be a three-state program as shown in Table 1. Stage 1 is not a formal program but rather the combined activity of the space development community (both inside and outside government). It includes activities that (1) contribute to space industry, (2) can be justified on their own economic merit and therefore funded by whatever means are available, public or private, and (3) help convince policymakers to embrace SRSI. A strategy for Stage 1 activities is discussed toward the end of this paper. They contribute to space industry by maturing the necessary technologies, by establishing infrastructure in space that lowers the cost of operating in space (so then Stage 2 can be accomplished for less cost), by demonstrating to policymakers and the public the many benefits of space industry, and by building conviction among policymakers that the SRSI concept is feasible. The robotics revolution already occurring in terrestrial industry will also help show that SRSI is feasible. Ideally, Stage 2 would begin today. There is no reason to have Stage 1 except for the fact that Stage 2 is not yet funded, so we must take practical steps to help convince policymakers to begin Stage 2.

**Table 1. Three Stages of Space Industry**

| Stage | Economics | Leadership | Goals | Duration |
|---|---|---|---|---|
| 1 | Each activity justified on its own merits | Amorphous | 1. Make incremental progress in technology<br>2. Lower cost of operating in space<br>3. Convince policymakers to fund SRSI | Until leaders are convinced |
| 2 | Focused, Multi-decade investment | Coalition of Nations | Intentionally bootstrap SRSI | 2 to 4 decades |
| 3 | Payback of prior investments | Coalition of Nations | Benefit from SRSI | Perpetual |

A model showing how Stage 2 can be done affordably was presented by Metzger et al.s [1] and will be briefly summarize here. We now know that the Moon and Near Earth Asteroids have all the raw resources necessary for an industrial supply chain (e.g., lunar ice containing hydrogen, carbon, and nitrogen, and the regolith containing silicon, metals, and calcium) [4-10]. Space industry can robotically mine these bodies for robotic manufacturing, beginning by making those materials that are easiest and require the least infrastructure to produce. For example, it can mine water for propellant and make crude metal from lunar regolith for building structures. Over time the industry can expand by receiving additional hardware from Earth at the same time that portions of the hardware are being made in space. The industry works to broaden its capabilities until it makes all the materials and parts in space, having developed a complete supply chain. Stage 2 should only take a few decades given adequate funding.

Stage 3 begins when space industry no longer requires imports from Earth and it becomes sufficiently autonomous that it no longer requires teleoperators to control every machine, so it can be loosely supervised and directed to grow exponentially without much further expense. At that point it becomes capable of making things of great value, both to enable in-space objectives and to return benefit to Earth. Its exponential growth will be rapid so it can begin providing large-scale benefits by the middle of the century. When teleoperation is no longer so important, the majority of space industry can be relocated to the asteroid belt where the greater accessible resources of the inner solar system are located. It can then extend its support of human activity to the outer solar system, and it can support even larger objectives such as terraforming Mars [11,12].

Ethical objections have been raised against space development, but others are claiming it is an ethical imperative. James S.J. Schwartz [3] has considered these positions and argued that (1) space development does not have a strong moral force behind it because it is unlikely to improve human welfare except in the far future, and (2) it will disrupt the efforts of space science, which actually does have strong moral force behind it. Therefore, he concludes that government should not change the legal and regulatory environment in ways that would help space development succeed. I will argue that both of his points are incorrect and that helping space industry to succeed should be a high priority. Although there have always been instances of development and science conflicting, we should ask whether the primary relationship between economic development and science has been one of conflict of one of mutual support. I agree with some additional concerns that Schwartz raised about commercial space development. For example, if asteroids have trillions of US dollars of value, then commercial asteroid mining could increase the disparity between rich and poor. I will argue that there is a large and necessary role for government in the bootstrapping of space industry. Thus, this concern can be mitigated.

The paper is organized as follows. Section 2 will discuss how robotic technology is changing the prognosis for starting a supply chain in space. It will argue that the technological barriers to SRSI are in the process of falling away and that there is no reason to delay Stage 2. Section 3 develops a rough cost estimate. It argues that the cost is low so again there is no reason to delay Stage 2. Section 4 discusses the scientific benefits of space development. It also considers how robotics technology is changing the discussion about humans performing space science missions versus just robots alone. I argue that when space development is considered there is a very strong argument for humans. Section 5 will discuss the potential economic, environmental, and existential benefits of SRSI. Section 6 will discuss a strategy for Stage 1 to convince policymakers to embrace SRSI. Section 7 is a summary with conclusions.

## 2. Technological Readiness for Space Industry

By the end of Stage 2, robots must be capable of all mining, manufacturing, construction, and associated tasks with minimal intervention by human teleoperators. With inadequate autonomy, the industry will be constrained by the cost of human labor and cannot be scaled-up to provide the greatest of the envisioned benefits. Robotics is not sufficiently advanced yet, and this may be seen by some as a large risk indicating that we should not start the project. However, robotics is already more than adequate for the early stages of space industry, and that will provide immediate scientific benefit as discussed in Section 4 while improving the cost effectiveness of space exploration. The robots can start as fully teleoperated from Earth[1] while they are few in number. Even before going to the Moon, each new piece of industrial hardware will be proven on Earth through field tests at lunar analog sites. Once on the Moon, they can be operated in the presence of a lunar outpost so astronauts provide additional troubleshooting and insight. Humans are not strictly required but will make it easier and faster to start the industry, which is important because the benefits will be obtained sooner. Furthermore, we have scientific reasons to put astronauts on the Moon [13], and the nascent industry will provision the outpost with oxygen, water, and spare parts, making it affordable and increasing its scientific value. As the lunar supply chain is bootstrapped and diversified, robotic advancements discussed below will allow us to increase the ratio of robots to human teleoperators. This trend will make the labor cost stay reasonable throughout the process. Within a few decades as robotic artificial intelligence matures, the teleoperators can adopt a loose supervisory role over the robots and the industry can grow exponentially. There is nothing to suggest we should wait for robotic autonomy to improve before we start the project.

According to Bill Gates in 2007, robotics "is developing in much the same way that the computer business did 20 years ago" [14]. Terrestrial industry has been slow to adopt automation, so the factories on Earth do not give a true picture of what is already possible. For example, less than 8 percent of transportation tasks in the U.S. are automated although 53% could be right now [15]. Widespread adoption of automation is expected to begin this year because cost is dropping steadily even while capabilities are increasing, including the interpretation of 3-dimensional vision and ability to work in unstructured environments. A 2016 review finds "robots dexterous enough to thread a needle and sensitive enough to work alongside humans. They can assemble circuits and pack boxes. We are at the cusp of the industrial-robot revolution" [16]. The sectors most needed for space industry are also the most automatable: manufacture of computers and electronics, electrical equipment, components, appliances, transportation equipment, and machinery are at least 85% automatable today [17]. Sectors that are not as important for the early stages are the ones that need more progress, such as wood products, paper, food, and textiles.

---

[1] The 2.7 s or less round-trip communication delay is manageable for many tasks, but robots will need some degree of automation for fine tasks.

Not just manufacturing, but other industries are converting to robotics. Refineries and chemical plants have long used automation in their process control systems. Mining companies are now adopting fully autonomous vehicles for underground mining, dozing, and hauling [18,19]. Robotics is penetrating the construction industry, as well. Masonry buildings are constructed by robots that lay bricks three times as fast as humans [20]. Robotic systems are being developed to autonomously construct steel beam buildings, including high-rise buildings [21].

Other than maturing the robotics, the technological progress most needed for space industry is to adapt existing technologies to the space and planetary environments. This is already on-going, including mining vehicles [22,23], low-gravity regolith conveyance systems [24,25], beneficiation systems to concentrate minerals found in lunar regolith [26,27] and chemical processors to extract resources from the minerals [28-30]. Their maturity level is adequate for the beginning stages of industry, and the point of bootstrapping is to develop the technology as we go along. Therefore, we can begin now.

Integrating the many machines and activities into factories will be challenging, but systems engineering methods were developed to handle this and they have a long history of success. An example is found in this fully automated factory in Japan:

> At this moment, in one of Fanuc's 40,000-square-foot factories near Mt. Fuji, robots are building other robots at a rate of about 50 per 24-hour shift and can run unsupervised for as long as 30 days at a time. When they stop, it's because there's no room to store the goods. Trucks haul off the new robots, the lights are cut, and the process begins anew. "Not only is it lights-out," says Fanuc vice president Gary Zywiol, "we turn off the air conditioning and heat too." [31]

Since then, more "lights out" operations with only robots, no humans, have been established including machine shops, distribution centers, and factories making things as diverse as food and electronics [32,33]. The extensiveness of automation is described for a 2015 Chinese factory that makes cell phone modules:

> …all the processes are operated by computer-controlled robots, computer numerical control machining equipment, unmanned transport trucks and automated warehouse equipment. The technical staff just sits at the computer and monitors through a central control system. [34]

Integrating the machines is becoming less complicated as robotic intelligence improves so the work environment can be left unstructured and the robots will figure out how to interact on their own. A helpful way to discuss it is by referring to Hans Moravec's four phases of robotic intelligence [35]. These correspond to (1) basic sensory perception and manipulation tasks ("lizard-like" robotics), (2) learning on its own through experiences in the real world ("mouse-like"), (3) maintaining a mental model of the world so decisions can be based on assessment of outcomes before doing them ("monkey-like"), and (4) general intelligence including the power of abstraction ("human-like"). In factories with thoroughly structured work environments like the lights-out factories that already exist, only lizard-like robotics is necessary. However, already by 2011 inexpensive workstation microprocessors [36] surpassed the 100 Giga Instructions per Second (GIPS) mark in computing power that Moravec estimated is necessary for mouse-level intelligence.

The software to take advantage of this hardware power is also making great strides. "Lizard-like" perception capabilities are routinely designed into robots using a wide variety of sensor systems, enabling them to perform more complex tasks in unstructured environments and even to interpret human intention [37,38]. Mouse-like learning is now becoming reality, as well. Deep learning (where layers of simulated neurons are trained in a process mimicking human learning) is having remarkable success [39]. Platforms like Facebook use facial recognition developed through Deep Learning techniques, with error rates dropping by 25% to 50% year-by-year so it is now 96.4% accurate [40]. While factory robots in space do not need facial recognition, the point is that deep learning works as a method and eliminates the need to design software algorithms for complex tasks. Factory robots in unstructured environments will need to identify objects they encounter, and programming that capability would be a terribly difficult task, if not impossible; deep learning has already enabled robots to visually identify objects from among 1,000 categories, and progress is ongoing to identify objects from among 10,000 categories [41]. Deep Learning also improves efficiency of factory tasks. It is enabling robotic welders to continuously improve by watching their welds through cameras [42]. Other factory robots are teaching themselves new tasks such as picking objects they have never seen before out of a bin. A robot is told generally what to do without being told how, and after one night of self-guided practice it can perform the task as well as if an expert had programmed it [43]. Then, the factory robot instantly transmits what was learned into the neural networks of all the other robots in the factory [43,44]. The sudden progress in robotic learning suggests we may have mature mouse-like robotics before 2030, the date Moravec predicted [45], so designing the SRSI machines to work together will not be so difficult.

3. **Cost of Space Industry**

Technology is not a barrier to beginning Stage 2, so next we consider whether there is a cost barrier. Unfortunately only a few rudimentary studies of SRSI have been performed and they are not adequately detailed to inform a good cost estimate, but we must bound the cost as well as we can using those studies.

The first study [46-48] was done in 1980 and concluded that 100 tons of factory hardware needed to be sent to the Moon to initiate SRSI. That was for only 80% closure (meaning 20% of the mass of a newly fabricated factory would be sent from Earth in each reproduction cycle), so a factory with 100% closure would need more diversification of manufacturing processes and thus would be more massive. The concept has also been studied for starships that travel to other solar systems then replicate [49], so 100% closure would be required, and a 2004 study considered it again for the Moon [50]. Unfortunately, landing 100 tons or more of hardware on the Moon is expensive, and the troubleshooting to make it work all at once would be overwhelming, and the technologies are not available yet for a complete supply chain in space. If we have to do it all at once, then we cannot get started. This is where the concept of bootstrapping comes in: instead of launching 100 tons of factory hardware, we can build it up over time using mass that is already in space. Similarly, when settlers moved to a new continent, they did not carry entire factories from their homeland across the ocean. Instead, they built up native industry over time utilizing local materials. Metzger, et al. [1] modeled this strategy to see if it can result in a net reduction of launch mass. The model suggested that it can: only 12 tons of assets and parts landed on the Moon over the span of few decades resulted in 150 tons of factory hardware at completion. Thus, it suggested a better than 10:1 reduction in launch mass might be achieved. The model also suggested that if 42 tons of assets and parts were landed, then the industry could have up to 100,000 tons of assets at the end of the same period, indicating better than 20,000:1 reduction; so it is non-linear. The modeling was simplistic and only intended to demonstrate the concept to motivate a larger study, which has not yet been funded. Bootstrapping also spreads the cost over a longer time, allows us to get started right away, matures the technology in the same environment where it must operate, and provides immediate scientific benefits by provisioning outposts or other activity in space. Since we were going to do many of those other activities anyway, they help defray the startup cost.

To make a very crude estimate of Stage 2, we note the cost of International Space Station (ISS) was about US $150 billion [51], and it consists of about 420 tons of hardware [52] so the rate is US $360 million per ton of human-rated systems delivered to Low Earth Orbit (LEO). The gear ratio to the lunar polar region is about 4 [53], meaning that for every ton to be landed on the Moon about 4 tons must be launched to LEO. Therefore, 12 to 42 tons to the Moon's surface requires about 48 to 168 tons to LEO. This suggests that if the hardware were given the same pedigree as human-rated systems it would cost between US $17 billion and US $60 billion. Spread over three decades, that is US $0.57 billion to US $2 billion per year, about 3% to 12% of NASA's budget, or 2% to 7% of the space budgets of the ISS partner nations. However, the hardware for space industry need not be human-rated, it will be based on existing terrestrial factory technology, and much of it will be copies of the same parts for many identical robots, so the development cost should be far less than this estimate predicts. Also, the launch costs using expendable rockets will be much less than the human-rated Space Shuttle. It is highly desirable to do the bootstrapping in the presence of a lunar outpost. Without that, the idealized model of Metzger et al. [1] may prove far too simplistic as hardware fails and there are no humans to troubleshoot and repair it. (This is discussed further in Section 4.3.) It is difficult to constrain how far off that model could be, but instead I will assume this project does take place at a lunar outpost. Spudis and Lavoie [54] studied a lunar ice mining outpost that supports two to four crewmembers and utilizes the ice resources to lower its own cost. The study included the development of all transportation vehicles except launch vehicles from Earth. They estimated[2] the total cost of the program would be US $88 billion in 2010 dollars, with a peak annual cost of $6.65 billion in year 13 of the 16 year program. Assuming worst-case that this peak level is constant and perpetual until SRSI is established so the human presence can grow as industry grows, then the cost of an outpost would be about 22% of the civil space budgets of the ISS partner nations for a total cost (outpost plus industry hardware development) of less than 30% of the combined civil space budget.

---

[2] They used NASA's accounting method then doubled the entire estimate [Paul Spudis, personal communication].

The costs are very low compared to the benefits. The primary barrier is neither technical nor economic, but just the feeling that it is a "nutty fantasy", so this is why we are in Stage 1 for now. A cost estimate of the possible activities in Stage 1 is beyond the scope of this paper, but the idea is that each step is commercially profitable and/or makes the national space programs more cost-effective so the space agencies will fund them from existing budgets. Stage 2 is where the intentional bootstrapping of SRSI begins, requiring sustained investment for several decades as estimated above. The investment toward SRSI beyond the cost of the outpost alone will produce increasing benefits for the outpost, including the ability to manufacture spare parts, build new experiments or tools, and construct radiation shields over habitats. However, it may not pay back the entire marginal investment until Stage 3, defined as the point where space industry becomes self-sustaining. It is beyond the scope of this paper to estimate the economic value of those benefits because a larger study is required. However, the following sections will generally describe some of the benefits in Stage 2 and 3.

## 4. Scientific Benefits of Space Development and SRSI

*4.1 Direct Benefits for Space Science*

Other articles in this special issue discuss specific ways that mining and manufacturing with space resources will make space science more effective. These include making rocket propellant from asteroidal or lunar volatiles, building radiation shields and landing pads from regolith, providing breathing oxygen and water for outposts on the Moon or Mars, and processing regolith to make metal for 3D printing spare parts. These activities will make space missions more effective because they reduce the amount of mass that must be launched from Earth while providing more flexibility during missions to respond to problems. The ultimate fulfillment of this is to build SRSI so that only humans need to be launched from Earth, something Mason Peck (former NASA chief technologist) dubbed "Massless Exploration" [55]. The full supply chain could then build entire spacecraft and a transportation network for scientists to visit bodies throughout the solar system. It could build grand astronomical observatories in space like the one described by Seth Shostak consisting of many electronically linked telescopes orbiting the sun, which could see a small automobile on the surface of an exoplanet 100 light years away [56]. It could build a planetary-sized particle collider on Mercury driven by solar energy to probe physics at levels we cannot currently dream of exploring, which might bring radical discoveries that benefit our civilization in ways we still cannot guess. Since SRSI can provide much greater leveraging of human labor than our existing economy, it can support any scientific activity (or artistic, literary, or other activity) we wish to explore by provisioning an institute located somewhere in space.

*4.2 General Relationship of Science and Economic Development*

Schwartz [3] raised the concern that space development will disrupt space science and therefore (since he believes it will provide little benefit to humanity except in the distant future) we should not be eager to get it started. He wrote:

> Let me first repeat that I have no in-principle objection to space development; my concerns arise when space development risks conflicting with space science…Though I grant that few conflicts exist at present, as space development and science capabilities improve, so too will the chances that development and science activities interfere with one another….
>
> There is much viable science that can be conducted only prior to development—that is, before settlement or exploitation contaminates, disturbs, or destroys sites of scientific interest. [3]

The problem with this objection is that it assumes we were actually going to do science at these asteroids or lunar sites before development could destroy or disturb them. If we were never going to investigate them anyway, then the argument does not work. Since 1991, humanity has sent spacecraft past 12 asteroids, about 2 years per visit. There are over a million asteroids larger than 1 km in diameter [57], and several billion over 10 m in diameter [58]. At the current rate of scientific investigation, we could visit them all in about 10 billion years, except the sun will not last that long. Mining will cause us to see far more than we would, otherwise. The same argument can be made for lunar mining, which may be largely underground to escape the radiation problem, and therefore will reach materials that would not otherwise be accessible for science, but which have immeasurable scientific value. This is also similar to our experience on Earth. Some subsurface exploration is funded for purely scientific reasons, but vastly more is funded for economic reasons, and the influx of data from economic geology has been a tremendous boon for science. The U.S. Bureau of Labor Statistics database shows that far more geologists are funded by economic activities than by pure science.[3] We can expect similar trends in space. Economic planetary geology will fund thousands more scientists who will publish papers on the great influx of data coming from space mining missions, including samples from the mined asteroids and the Moon, telescope observations for prospecting, and theoretical modeling to understand which asteroids or lunar locations are better for mining.

Furthermore, scientific productivity is correlated to economic development. Gantman [59] studied the publishing record of scientists in 147 countries and found that scientific productivity correlates strongly (with high statistical significance) with the size of the country's economy measured by Gross Domestic

---

[3] It shows that 65% of geoscientists are working in mining and other economic fields, 18% in research (some of which is economic research and some is pure science), 12% in government (mostly managing economic activities), and only 5% in academia (some of that is teaching and some is scientific research).

Product (GDP) and with the development of the country measured by GDP per capita. The idea is that large economies have more money at their command to direct into scientific research, and developed countries afford scientists better access to the scientific equipment and infrastructure needed to perform the research [59]. The mechanism that makes this true country-by-country should also make it true for the world as a whole, as well as for the "country" we call space. As the global economy grows, more funding is available for science, and as civilization becomes developed, scientists have greater tools and other resources at their disposal. When SRSI leverages human labor by another factor of 10 or more, we can expect GDP to increase so both funding of science and access to resources for science will improve. This will enable scientists to make even better use of the vast, new data sets coming back from the increased activity in space.

As argued above, the initiation of SRSI (Stage 2) can be accomplished for a minority fraction of the existing national budgets for space science and exploration. Therefore, if we were to gain no benefits from SRSI other than the scientific ones, it would still be worthwhile. The benefits to science will accrue not only in Stage 3, but also in Stages 1 and 2.

*4.3 Relationship of Human and Robotic Space Science*

The advances in robotics technology also change the discussion whether science and exploration should include human-tended (crewed) missions or only robotic missions. As robotics are becoming ever more capable, it is possible that a single robot may one day (perhaps soon) produce as much value per hour on a planetary surface as a human does while remaining far less expensive. This might intensify the questioning why we have humans doing science and exploration in space.

On the other hand, for space development activities there is a very strong argument for human spaceflight. At present, robots are not sufficiently advanced for troubleshooting and repairing other robots. Without humans present, when a mining or manufacturing robot breaks, operations at the manufacturing site may be stopped until a replacement robot is sent from Earth. These failures can be expected frequently because terrestrial technology in these fields has not yet been well-adapted to the lunar environment, and because these activities require greater ruggedness than robots just scientifically sampling. Furthermore, we will probably have difficulty identifying the root cause of hardware failures or operational problems on the Moon, so without humans on-site the replacement robots may exhibit the same or related failures multiple times before an adequate diagnosis is made. Humans are uniquely capable of performing these tasks at a lunar outpost. Thus, with humans present, advances in space development can be far more rapid than by purely robotic space development activities alone. This means the great scientific benefits of space development, as well as the economic and environmental benefits discussed in the next section, will begin to accrue much sooner. Because SRSI in Stage 3 grows exponentially, the benefits will also accrue exponentially. The difference between an exponential curve and a delayed exponential curve diverges to infinity. This suggests that the human contribution in space development missions is an immeasurable value per dollar cost. This model is too simple but it illustrates the point that value compounds geometrically in development missions, which changes the calculus for human spaceflight. Knowing how SRSI may help solve the world's economic and environmental problems, the public may have much stronger support for a lunar outpost if the astronauts spend a portion of their time standing up lunar industry.

Human presence in space is also made more economical in a space development scenario and this contributes to the reasonableness of human spaceflight. As Spudis and Lavoie have demonstrated [54], a lunar outpost based upon the use of lunar resources is less expensive than one in which everything is launched from Earth. Robots and humans will be highly synergistic in Stage 2, and in Stage 3 the robots will build spacecraft and outposts that open the solar system for humanity.

## 5. Economic, Environmental and Existential Benefits

The economic, environmental, and existential benefits are more speculative since we cannot tell what will change several decades in the future. I wish to emphasize that we have strong justification for space development whether or not we are sure to gain these additional benefits. There are at least two reasons for this. First, space development is already justified by the scientific benefits. Since we are already spending money in space, and since the existing budgets are adequate, then the most effective use of those funds should be pursued, and that includes space development. Second, the humanitarian and environmental challenges discussed below do not have any completely sure solutions, either terrestrial or space-based. As long as space development provides at least some significant, additional chance of solving them, then there is an ethical burden to pursue that chance.

*5.1 Economic and Environmental Benefits*

Economic and environmental benefits are linked and will be assessed together. The benefits may include materials and manufactured goods sent to the Earth from space, space-based communications systems and computing linked to Earth by those communication systems, and space-based power beamed to Earth.

5.1.1 Minerals and Manufacturing

Civilization and the industry that drives it are stressing the planet in a number of ways, including exhaustion of the most enriched deposits of some economically important minerals [60-64]. This will be exacerbated by our economy expanding to provide developed economies to all 7 billion people, soon to be 11 billion [65,66]. Some of these minerals are important to renewable energy and the shortage threatens our ability to scale-up renewable energy to global levels [67-69] Mining in space can help replace these minerals, making renewable energy (and other important technologies) more affordable [70-72]. Some manufacturing processes will also benefit from the space environment [73]. These benefits will begin early in space development, and as Stage 2 progresses into Stage 3 then it may become economic to bring a wider range of space-made goods and materials back to Earth.

5.1.2 Data and Computing

Exponential growth is occurring in computing and the internet, and since it is exponential it can quickly reach physical limits. When communication channels reach their capacity then more spatial diversity is needed to transmit data in parallel. For example, more fibers must be laid across the oceans, or satellites must use narrower antenna beams so they can transmit to smaller spots on the ground side-by-side. Installing spatial diversity is expensive and becomes the limiting factor when costs are too high or no more diversity is possible. A recent meeting at the Royal Society concluded that a number of "capacity crunches" are looming [74]. The world population is increasing, the percentage using the internet is growing [75], 50 to 100 billion physical objects (in addition to the people) will be using the internet by 2020 [76], and individual users' data rates are growing exponentially with a 50% increase per year for thirty years straight [77], which is analogous to Moore's Law for transistor density. A 2015 report by the Semiconductor Industry Association and the Semiconductor Research Corporation says, "Unfortunately, neither existing technologies nor current deployment models will be able to support the skyrocketing demand for communication, especially in the wireless sector" [78].

Another issue is with the energy that computing uses. The semiconductor report [78] also estimates the growth of computing in terms of binary bit transitions per year and the energy used for each one, including a physics-based theoretical lower limit, and this shows that if this growth were to continue then by 2040 our computing and the Internet would consume all the energy the world produces. That is only the operating energy for computers; the energy to manufacture them is far more, since for example the energy to manufacture just one memory chip exceeds the total energy to operate a laptop computer for three years [79]. Considering the manufacturing needed to support the growth rate of data storage and data transmission [78], global energy limits could stifle progress in computing well before 2040.

One possible solution is to offload the burden to space. Presently, data communication between Earth and space is far too slow for satellites to handle much of the internet's traffic or to locate much of our computing off-world. This can change with space industry. First, SRSI can build large phased antenna arrays in orbit, large enough that beam widths can be very narrow. I calculate that for the antenna's Fraunhofer region (far field) to begin at the distance to the Earth's surface from geostationary orbit when operating at 38 GHz, the antenna would be 375 m in diameter and the beam spot would be 1 km in diameter on the ground. Since the United Kingdom has a land area of 243,610 km$^2$, about 243,610 beam spots would fit within the UK alone. This is about 4.6 times more than the number of mobile phone base stations (cells) presently in the UK [80] Power densities at the ground can be kept lower than mobile base stations as needed for biological health, and antenna arrays on Earth can still receive the signals. With a Shannon Limit [81] of about $10^{11}$ or $10^{12}$ bits/s per beam, this would provide a data rate of 25 to 250 petabits/s for the UK, which is 20,000 to 200,000 times the entire average Internet rate of the UK in 2015 [82]. Spatial diversity can be increased further because these beam widths are a tiny fraction of a degree, so a single ground station with a phased array can receive channels from thousands of points around geostationary orbit, while each point in geostationary orbit transmits thousands of different channels simultaneously, each to a different point on the ground. Thus, microwaves alone have the potential to extend the UK's Internet by a factor of 20 million to 200 million when we can build vastly larger antennas in orbit. Furthermore, SRSI could manufacture laser optical communication systems for even smaller beam spots, tighter directionality, and higher data rates per channel, although they would not work through cloud cover. The combination of both optical links and microwaves (to supplement during cloudy conditions) across the globe, with terrestrial fiber optics routing data that was diverted due to cloud cover, could vastly exceed today's global Internet throughput.

With such a communication network, the bulk of computing can be located off-world, relieving the Earth of the energy demands for both manufacturing and operating computers. Only the compressed data streams for user interfaces need be transmitted to and from the Earth. While space communications will have greater latency, there are techniques in conjunction with terrestrial networks and computing plus increasing artificial intelligence to make it acceptable. Since the in-space computers and network will be made by the autonomous SRSI, the cost will be very small. This will give humanity the benefit of much more computing than terrestrial energy and resources could ever provide, escaping the exponential limits of growth. This may be extremely important to our future.

5.1.3 Energy

Space industry could also move power generation off-Earth by building a Space-Based Solar Power system (SBSP) [83-86]. There is widespread belief, which I share, that it is presently too costly to launch SBSP from Earth to make it worthwhile [87] except for special applications where cost is not an issue. Solar cells located on Earth's surface with batteries (to extend solar energy into the nighttime and cloudy days) can provide the same service more cheaply than a system launched from Earth into space. However, if we have SRSI, then building SBSP will be low cost since it will come from excess manufacturing developed through self-replication. Analogously, hunter-gatherer societies do not pay for the food chain to grow their food or building materials, because the food chain is self-replicating and autonomous from the human perspective. SRSI can be considered a designer food chain adapted to space. SBSP can beam carbon-free energy to Earth at far less than the cost of petroleum-generated power and far less than Earth-based solar energy with storage, so the predicted population of 11 billion people by 2100 [65,66] can live in developed economies without further environmental impact. With adequate energy we can solve other planet-scale problems, such as the shortage of fresh water, because energy can drive desalination plants, and we can begin reversing environmental damage.

Presently, if all humans worked as hard as agricultural laborers (about 0.1 horsepower = 0.074 kW [88]) then the 155.6 million laborers in the U.S. with 2000-hour work-years would annually produce 23.4 TWh (0.084 exajoules, or EJ) of thermodynamic work toward goods and services using their muscles. That compares to 28,600 TWh (103 EJ)[4] [89] of energy used annually by machines in the U.S. including oil, coal, gas, solar, nuclear, etc.[5] This says that machines do greater than 1000 times more thermodynamic work than human muscles, so less than 0.1% of things made or services provided in the U.S. are actually through human labor, while more than 99.9% are by machines. The role of humans has largely evolved into controlling machines and making decisions. This roughly 1000-to-1 leveraging in developed nations and 300-to-1 in the global average is what makes possible the diverse supply chain, which in turn enables the existence of modern technology. Our ancestors would have considered smart phones and jet planes to be just as fantastical as O'Neill's space settlement was for Senator Proxmire. Sir Isaac Newton could not have built a smart phone even if he had detailed instructions because his world had not bootstrapped the extensive supply chain required for it, including energy sources that give us the productivity to operate that supply chain. SRSI can reproduce up to the limits of resources in the solar system, which are billions of times greater than the resources on Earth [90], without harming the ecosphere so it can increase our leveraging by a factor of 10, a thousand, a million, or more, broadening the supply chain, enabling further technological progress, and taking human civilization to a higher level.

---

[4] 2015 total, reported as 97.523 Quadrillion Btu.

*5.2 Existential Benefits*

We cannot realistically deal with the existential threats of the solar system and galaxy without more leveraging than we presently have. With SRSI, we can begin seriously reducing those threats. SRSI can build enough spacecraft to survey all the small bodies of our solar system and build more spacecraft to travel to large numbers of them and adjust their orbits. It can put observatories throughout the solar system looking for dangers from beyond, and build robotic re-boot systems including genetic banks to restart life and civilization if disaster does strike. Microwaves from SBSP could power the world through the ash cloud of a supervolcano eruption when wind and solar energy sources would fail (causing more deaths than the eruption itself), and it could rapidly grow and manufacture relief supplies in space and send robotics to the surface to ramp up factory agriculture. Many more such examples can be imagined.

We could also rapidly terraform Mars [11]. First, autonomous robotic labor can capture icy bodies in the outer solar system and transport them to enter the Martian atmosphere. Second, it can build SBSP in orbit around Mars, and it can build Mars landing spacecraft bringing materials and equipment to provision the initial factories on the surface of Mars. This may require SRSI to transport material from the Main Asteroid Belt. Third, powered by SBSP, the Martian surface industry can eventually become self-sustaining then self-replicate to grow exponentially, produce greenhouse gases, remove soil salts, and rapidly transform the planet. Fourth, the Martian surface industry can then metamorphically recycle most of itself and move underground to leave Mars beautiful. SBSP will continue to support Martian civilization. When Mars can support a large population with no reliance on Earth, then we are truly a multi-planet species. That will provide buffering against disasters that happen on only one of the two planets. Simply putting colonies on Mars will not achieve this, since the colonies will be dependent on Earth and Mars will remain a very inhospitable environment. Only scaled-up SRSI holds the potential to make us a two-planet species. A study should be performed to quantify how quickly this could be accomplished, but considering that SRSI could exceed a million times the economic activity of all Earth within this century, it may not be such a distant goal.

---

[5] Much of that energy is not productive thermodynamic work because of theoretical efficiency limits and waste, but also the majority of U.S. workers use their muscles for productive thermodynamic work far less than agricultural workers. Therefore, this rough comparison is still valid.

Finally, it can take humanity and other life beyond our solar system. It could initially build robotic ships to take SRSI-bootstrappers to other star systems where the robots can locate and terraform candidate worlds in advance of our arrival [91-93]. It can then bring humanity to those worlds by building multi-generational ships for us [94-97], or it can take genetic banks (embryo ships) to slowly populate those terraformed worlds [98,99].

These ideas are fantastical today because our supply chain is too narrow and our industrial leveraging far too small. SRSI will change that. Without SRSI, we have no reasonable possibility of significantly reducing these existential risks. If this were the only benefit to SRSI, we should still do it.

One additional benefit of SRSI that relates to existential threats is the possibility that artificial intelligence and robotics could destroy humanity. One possible approach to dealing with this is to perform research in these fields at a safe location off the Earth. SRSI can enable this by building the computer systems off-world at a chosen location. This is not a complete solution, but it may provide important tools to move forward in this field with greater caution. It should be noted that SRSI itself does not require general artificial intelligence; it only requires the level of automation that we are putting into factories on Earth today to perform simple tasks, and therefore it is not a threat by itself.

*5.3 Objections*

Before looking at objections, I should clarify that am not claiming it is a provable fact that we can realize these economic and environmental benefits quickly enough to solve our present challenges. Instead, I am claiming there is a significant probability that space industry will help solve those challenges quickly enough, and it may turn out to be the only adequate solution, but if we do not attempt this solution then there is zero probability it will help. We cannot know for sure because the global challenges are complex. Also, I am not claiming the justification for space industry based upon these economic and environmental benefits needs to stand alone to be a valid reason to begin space industry (although I, personally, do believe they stand alone as valid). What I am arguing here is that these benefits provide supporting justification for space development when considered alongside the benefits to space science. Space science is already supported globally at funding levels that are adequate to perform space development, and since space development will make space science more effective and therefore give the public a better return on its investment, any additional benefits we might gain from space development are serendipitous. The fact that this serendipity just might save civilization, with no downside for trying, makes it an easy sell. With this in mind we can address the following objections.

5.3.1 Objection: Terrestrial Solutions Are More Likely

One objection is to say that space-based solutions are not necessary because terrestrial technologies, especially nuclear fusion and renewables such as solar, wind, and geothermal, coupled with improvements in efficiency and sustainability, will probably solve Earth's energy problems. The weakness in this objection is that there are significant unanswered questions about these other technologies, so we do not know for sure they will provide solutions that are adequate or come quickly enough although the stakes are high. Furthermore, it is not an either/or decision. The global energy system may be most effective when it includes the synergy of terrestrial and space-based solutions together. One example is space mining that could provide necessary but increasingly scarce elements for renewable energy technology. A few of the specific questions about nuclear fusion and renewable energy are reviewed below along with ways that space-based solutions provide synergistic or alternative solutions.

One concern with nuclear fusion is that the world must transition away from fossil fuels as quickly as possible, both because more affordable fuels are expected to decline this century [100,101] and also because they add carbon to the atmosphere [102], but nuclear fusion might take too long to develop. The timeline of the world's leading fusion effort, consisting of the International Thermonuclear Experimental Reactor (ITER) followed by a demonstration reactor (DEMO), was planned for completion about 2050. An economic model based on historic trends says there will likely be three demonstration reactors in the early 2050s, ten operational reactors in the early 2060s, and 100 improved reactors in the 2070s, which will still provide only a small fraction of the world's energy at that late date [103]. The study says that after a brief period of exponential growth it enters into a linear growth phase. We would need to build 200 reactors *per year* in that phase to reach 30% share of the world's energy by 2100, and that assumes no further delays. In 2006 the plan for ITER was to achieve first plasma in 2016 [104], but it is now planned for late 2025 [105], slipping 9 years in 10 years, almost year-per-year, and the previous planning was declared "totally unrealistic"[106]. Considering the significant knowledge and technology gaps that remain [107-110], and that we do not know what we do not know, there is a risk we will again find our plans have been unrealistic. Some are even arguing that economic fusion is not a reasonable goal [111]. Space-based solutions can provide global scale energy on the same timeline or even sooner than fusion with arguably less schedule risk since no new physics are required. Because SRSI does not depend on the world's limited resources or labor force, it can scale-up exponentially while terrestrial energy solutions like fusion can only scale up linearly. Pursuing both paths simultaneously will give humanity the lowest total risk.

A second concern about nuclear fusion is that it may not be sustainable when scaled-up to global power levels. The fuels are sufficient to last millions of years, but some other elements needed to build and operate the reactors may not be [112], so we must develop alternative technologies that do not need those scarce elements (entailing additional technical and schedule risk beyond ITER and DEMO) or we must develop new, extreme extraction methods to obtain those elements (entailing new technical risk, higher cost, and further possible damage to the environment [112]). The concern is that eventually "the cost of producing a further ton in terms of energy, water, and environmental damage will be so great, that mining will cease" [113]. Elemental scarcities may be relieved by space industry that mines asteroids and returns the "vitamin" elements back to Earth, thereby making fusion more affordable with less environmental damage, so there can be synergy between the two approaches. Also, after we have space industry then SBSP may be a less expensive energy supply in terms of ongoing costs, so fusion might be limited to more appropriate niche applications and the global supply of those elements will not become so stressed.

A third concern with fusion is that it is not perfectly safe [114] or environmentally friendly [115], so although it is vastly better than fission the politics could still turn negative. Although the risks are small, the risks from SBSP may be smaller since the microwave beam would have low power density [116,117].

A concern with switching from fossil fuels to the optimum mixture [118,119] of renewables like solar and wind (requiring a smart grid with storage and/or excess generation capacity, along with nuclear fission and/or fusion) is that it is expected to reduce the world's net energy return on energy invested (EROI) [120,121]. Roughly, EROI is the amount of energy that the energy sector produces per unit energy that it consumes. It is usually calculated for each energy source separately to compare their effectiveness. The EROI of petroleum has already dropped from greater than 100:1 in 1930 to less than 20:1 by 2005 [120] as the cheaper petroleum reserves have been progressively exhausted, and this may already be hindering the global economy [120, 122-124] including the economic growth of underdeveloped nations [125]. Some argue that lower EROI is not able to support the flourishing of modern civilization because it does not provide adequate leveraging of human labor to provide the full breadth of goods and services we have grown to expect [126-129]. Lambert et al. [126] found that measures of well-being, including the percent of children underweight, female literacy, and rural access to improved water, correlate to the EROI in a society, independent of the per capita energy expenditure [130]. Those data show that the EROI necessary for a higher standard of living is somewhere above 20 to 30, higher than terrestrial renewable energy is capable of providing. In fact, they conclude,

> Despite the best intentions of improving net energy balances for a developing nation, the large-scale introduction of renewable energy generation may have too low an EROI and may prove too expensive to facilitate continued growth in developing economies. [125]

SBSP can overcome this potential problem because it has an extremely high EROI (solar energy concentrated by an autonomous system and put into a form requiring minimal energy expense by humans to capture it). It can be deployed at global-scales or just locally in developing areas for little cost on the ground.

Another concern is that the public and its leaders may not invest adequately to set up the terrestrially-based renewable energy system in time to avoid economic and environmental disaster. A 2014 study estimates the U.S. will need to spend $600 billion per year for 50 years to build a smart grid with

adequate buffering for the intermittencies of renewable energy. The cost to build a renewable energy system may come down as advances in robotics make terrestrial manufacturing less expensive, but on the other hand a 2013 study of energy storage found the specific materials needed to build a global-scale system will require approximately the entire global annual production of those materials for many years [131]. For example, the study estimated it would take 28 years to mine enough lithium for batteries to support renewable energy at today's global power levels. If the world's power needs triple by 2100 as models suggest [132,133] then it should instead take 84 years of the world's mining to meet that need, but that fills the entire time. Since there are other economic needs for those same elements, the mining rate would need to effectively double for the rest of the century, increasing the cost of storage for renewables and making it even more likely the public will balk at the expense. The report finds that such (typically unaccounted for) costs will make a smart grid impractical unless energy storage life cycle can be increased by a factor of 5 to 10 [131]. Solving the energy problem with terrestrial renewables requires high levels of new spending, whereas bootstrapping SRSI can be done for far less completely within the existing space exploration budget, so it will be easier for policymakers to convince the public to make the investment. Furthermore, it is not an either/or choice. SRSI/SBSP can be developed in parallel with terrestrial systems for the best chance of success.

Penetration of renewable technologies into the global energy system may also be limited by environmental concerns at the energy collection sites [134], by available resources to manufacture the global system as mentioned above [135], and by the incremental cost versus the diminishing return of additional generators on the smart grid. Because of the intermittent nature of renewables, excess generation is required to get closer to full penetration, and the marginal addition of generation does not increase the power factor as much as the first generators on the grid; getting the next percent penetration is always more expensive than the previous percent [136]. Generators provide value at only 50% to 80% of their original value by the time penetration reaches as little as 15% [137]. One 2012 study concluded, "Research as well as policy should take the possibility of a limited role for solar and wind power into account and should not disregard other greenhouse gas mitigation options too early" [137]. Another study examining the resource limitations (which drive the costs for renewables) concluded, "an overview of the land and materials needed for large scale implementation show that many of the estimations found in the literature are hardly compatible with the rest of human activities" [135]. At present, although renewable energy generation is growing, market forces have not allowed it to make even a dent in the use of fossil fuels since energy demands are growing faster than renewable energy [120, 124]. Although penetration of renewables may reach no more than 10% to 35%, a 2016 study finds even that much carbon reduction will have a societal value of US $85—$1,230 billion (central value of $400 billion) [138]. Bootstrapping SRSI will result in 100% penetration and thus will provide much more societal value than it costs, not even considering the many other benefits.

In view of these significant questions, the wisest approach is to pursue all of the potential solutions. Friedrich Wagner, who headed several fusion reactor experiments, considered the dire need to replace fossil fuels and the potential for success in developing (1) fusion, (2) inherently safe fission, and (3) adequate energy storage so renewables may become baseload power, and he wrote,

> Facing these perils energy research has to be tremendously intensified. Only the results of energy research may be the insurance that 9 billion people on Earth can live in peace. Facing the few options mankind has and the doubts on their potential, acceptability and sustainability, none of the three possibilities may be discarded or delayed. [139]

Space development leading to SBSP gives us a fourth possibility to escape these perils, one that requires no new physics like fusion, costs less than renewables for higher penetration, and provides EROI potentially much higher than any other. It should not be discarded or delayed, either. Among all the possible solutions, SBSP is uniquely positioned to overcome the economic barriers facing renewable energy because it can already be justified from existing space exploration budgets. It is also uniquely positioned to overcome the resource and environmental barriers facing renewable energy because it is located outside the limitations of this planet.

5.3.2 Objection: Why Not Put the Self-Replicating Factories on Earth?

If solar energy has a low EROI when used on Earth, then how can a space-based solution have sufficient EROI when it is entirely driven by solar energy? It can for the same reason that a forest can collect solar energy and concentrate it as biomass. Energy that an autonomous system invests in itself through a net-positive metabolism, even if it is barely greater than 1:1 EROI, does not detract from human activity or lower our society's net EROI. This is why robotic autonomy is important for SRSI to be successful. However, if this autonomous leveraging gives solar a high EROI in space, then we might achieve the same solar EROI by bootstrapping robotic industry on Earth: a Self-sufficient Replicating Terrestrial Industry, or SRTI. After bootstrapping SRTI, it could build the renewable energy, storage and smart grid systems to concentrate energy so we can obtain it without much energy investment involving humans. Earth has many advantages over space that would make it easier and cheaper to start robotic industry, so what is the real advantage of going to space?

The answer to the objection is this: that we cannot scale up industry by a factor of a thousand – or even ten – here on the surface of Earth without causing great harm to the Earth's ecosystems. The primary benefit of space is real estate that biology does not need. Earth is the one special place in the solar system required by life, but machines can function anywhere else. We cannot scale up computing forever while it remains exclusively on Earth, and we cannot greatly increase mining of the necessary elements. Getting industry off the Earth's surface is the goal, not the means to the goal. All the global challenges we now face – population density, global warming, shortages of fresh water, cost of energy, running out of the cheapest ores so we must move on to mining the more expensive ones – are symptoms of a common problem: we are getting close to the limits of a planetary civilization[6]. Because of that limit we cannot forever continue doing greater things as a species while relying on a single planet.

5.3.3 Objection: Space Industry Cannot Help Enough

Another objection is that it is unreasonable to think we can put enough of our industry into space to make a difference. Lunar industry can make SBSP to beam energy to Earth as massless photons, but transporting manufactured goods that have mass is more problematic.[7] The energy sector is a small fraction of the overall economy (there are more machines using energy than machines generating it), so how can it be worthwhile to move only the energy sector into space?

I have already argued that computing can eventually be put into space after SRSI has been bootstrapped, and this will relieve a potentially giant burden from Earth's environment. In addition, moving the energy sector, or a part of it, off-Earth is not a negligible accomplishment. Projections say world population may exceed 11 billion by the end of the century and power demand could reach 51 TW (1610 EJ annually)[8], four times the current demand [132,133]. That means there will be (roughly) four times as many machines on Earth using all that energy. The EROI of petroleum is about 20:1, meaning for every 1 unit of energy running machines to obtain petroleum we get back 20 units of energy to power all the machines [140]. That means (very roughly) $1/20^{th}$ of the machines are working for the energy sector. As EROI declines both through exhaustion of petroleum and through adoption of lower EROI renewables, the fraction of machines supporting the energy sector should become much higher. Even if we can put no more than this portion of the machines on the Moon, that will reduce the burden on Earth's environment by a great amount.

However, space industry can do even better. Low cost, abundant energy will give us the ability to fund grand environmental projects. For example, over time, all of Earth's manufacturing could be moved underground where it will not compete with ecosystems on the surface. Robotic mining has already become advanced enough to do this. Then, terrestrial industry can be scaled-up with no environmental impact. Robots could also mine and recycle the landfills, remove carbon from the atmosphere, build greener cities, and restore habitats that were previously damaged. Two things we cannot put underground are (1) the renewable energy generators, and (2) the entire industry as it builds up to the level necessary to start going underground. Getting to that point would cause terrible damage to the Earth unless we use space-based power.[9]

---

[6] The ethical limit of a planet comes sooner than, and is more relevant than, the theoretical limit described by the Kardashev Level. Achieving a Kardashev Type 1 civilization without leaving the planet would require coating the entire globe with solar cells and thus driving most plants and animals to extinction.

[7] It actually could deliver manufactured goods to the surface. SRSI could build a constant stream of one-way, landing spacecraft to bring down the manufactured goods. The landing spacecraft could be recycled for their metal and electronics for terrestrial use.

[8] The exact value does not matter for this discussion. A 2007 study [133] reviewed the 133 economic models in the literature that were subsequent to a 2000 climate report, and it found the $25^{th}$ percentile prediction near 1100 EJ, the median near 1300 EJ, $75^{th}$ percentile near 1800 EJ, and $95^{th}$ percentile near 2550 EJ. Since global development is crucial to stabilizing the population, then failure to provide adequate clean energy to enable adequate economic development will exacerbate the problem. Arguably, engineering conservatism requires we minimize risk by planning to meet the energy levels in the higher end of the range plus a safety margin. 1610 EJ is only between the $50^{th}$ and $75^{th}$ percentiles and is therefore not too conservative, so it is not an exaggerated figure in this discussion.

*5.4 Weighing Probabilities*

There is significant uncertainty whether renewable energy will be adopted quickly enough and with sufficient penetration to avoid economic or environmental harm. There is also uncertainty whether it will provide adequate EROI for all nations to become highly developed and for civilization to be at its most vibrant. In my opinion SRSI/SBSP has the best chance of solving these problems because Stage 3 is not constrained by economics, global resources, or ecological impacts, and because the scaled-up robotic autonomy results in extremely high EROI. Even in Stage 2 it is buffered from economic stresses if the space program simply maintains current levels of funding for science and exploration. Thus, it uniquely decouples the world's energy and economy from the world's environment. Suppose SRSI and SBSP become available only after terrestrial renewables have scaled-up to maximum 35% penetration; then SBSP can still be a contributor by providing the other 65% to get us completely off fossil fuels. Suppose SBSP has a 50% chance of helping avert global economic and environmental disaster, and suppose the development of economic nuclear fusion also has a 50% chance. Since the two solutions are statistically independent, pursuing both simultaneously provides a 75% chance of averting disaster, reducing the odds of disaster by another 25% over just fusion alone, and we get that extra chance for free by making space developing part of our space science within the existing budget. The seriousness of these potential humanitarian and ecological disasters makes this argument strong, because avoiding disaster is priceless, and even 25% of priceless is priceless.

Regarding the existential concerns, Matheny [141] assumed humanity could survive as long as *homo erectus*, so an extinction event has the expectation value of erasing 1.6 billion life-years from present and future humanity. He used asteroid impact statistics [142,143] to show that if half of these potential extinction events are preventable asteroid impacts, then a US $20 billion space system to deflect asteroids has an expected cost of only US $2.50 per human life-year, far less than we spend per life-year on insurance and healthcare, to state the obvious. Bootstrapping full SRSI may cost US $9 billion per year for 40 years, so that is still only $45.00 per life-year and the cost is already covered by our existing space budgets. The value is even greater when we consider that it preserves all the other species on Earth, provides other remedies for existential risks and lesser emergencies, and provides other benefits for the environment, economy, internet, computing, and space science. Matheny concludes, "We take extraordinary measures to protect some endangered species from extinction. It might be reasonable to take extraordinary measures to protect humanity from the same." However, this measure is not even extraordinary since we can obtain it for free by adopting more effective space science.

---

[9] Even industry on the Moon can be underground and powered by space-based power collectors to minimize impact to the lunar surface. The power collection would be in halo orbits at the first and possibly second Earth-Moon Lagrange points.

Our situation in this generation is unique because we live during the historic rise in human population that produced the first planet-scale challenges, and we are the first for whom SRSI capabilities have become achievable. Therefore, we are uniquely positioned as well as uniquely responsible to deal with this. Our biological conditioning may be the reason it is easy to ignore future people we will never meet and events that will probably not occur in our lifetimes. It therefore might help the unconvinced to think how future generations will be grateful to us if we choose to make them safe and prosperous by bootstrapping SRSI, and how we can enjoy today the knowledge that we are doing it.

## 6. Practical Roadmap

Pragmatically (and unfortunately), Stage 1 seems to be a necessary stage to convince world leaders to embrace SRSI. If too much time is spent in Stage 1 convincing them, then it will delay the start of Stage 2 and the arrival of Stage 3, reducing the chances SRSI can provide help to the Earth quickly enough to avoid severe economic or environmental disaster. Therefore, we should adopt strategies to get through Stage 1 by convincing policymakers as quickly as possible.

*6.1 Practical Demonstration Required*

Some in the space settlement community believe the traditional, governmental leadership in space is underperforming, so they have pinned their hopes on newer, smaller commercial space companies to lead the way [144]. I disagree: while I believe commercial entities will make a substantial contribution, I do not believe they can do enough. When you examine a list of all the possible ways to make a profit in space, the activities fall short of completely bootstrapping SRSI. It may take some decades before "ignition" where the entire system pays back more than the ongoing investment. That is too long for individuals hoping to benefit from their investment, so they will invest elsewhere. This is why government must play the key role in leading its development, which brings us back to the original problem that government leaders must be convinced to invest in space settlement, something that has never happened before because it seems like a "nutty fantasy".

Nothing beats a practical demonstration for convincing people that something is feasible. The strategy for convincing policymakers (and the public they respond to) must involve smaller steps of implementing pieces of space industry, doing what we can as funding becomes available. The process of convincing leaders is therefore a process of actually starting space industry, but more slowly and piecemeal with inadequate funding, instead of bootstrapping it intentionally and as quickly as possible with adequate funding. This slower process delays the benefits of space industry, but as soon as the piecemeal progress has convinced enough of the public and their leaders that the overall program is achievable, and as the benefits become apparent, including its potential to solve world energy and climate problems, then I believe it is likely they will embrace Stage 2. This is especially true because Stage 2 already has enough funding if we simply direct the national space programs toward this purpose. The strategy to convince government leaders is therefore a matter of finding activities that:

> (1) can be justified within existing budgets and/or make a commercial profit,
>
> (2) contribute practically to the technology or infrastructure of space industry, and
>
> (3) demonstrate that space industry provides benefits.

There is an on-going discussion of strategies in the space development community about how we can convince government leaders to support space development. Some degree of consensus has emerged that the following steps are the most important:

- Space agencies will fund development of In Situ Resource Utilization (ISRU) technologies (e.g., water mining [25], oxygen extraction from regolith [28], and 3D printing landing pads from regolith [145,146]) to support their sortie and outpost missions. These technologies lower the mission costs and pay back the agencies in the near term. These technologies are a subset of what is needed for space industry, so developing them is also helpful for space development.
- The space agencies should eventually establish a propellant depot in space and give contracts to the space mining companies to extract water from asteroids and the Moon to make rocket propellant [147-151]. The propellant makes government space missions less expensive. These contracts give the mining companies cash flow, develop infrastructure in space that contributes to space industry, perfect the mining techniques, and prepare the mining companies for non-government customers.
- Space mining companies will begin selling propellant to space tug operators (or operating tugs themselves) to boost telecommunication satellites from Geosynchronous Transfer Orbit (GTO) or even from Low Earth Orbit (LEO) into Geostationary orbit (GEO) [147-149,151]. This will be a profitable revenue stream enabling space development companies to make further progress.
- The availability of cheap propellant will transform the goals of national space agencies, enabling them to do more frequent and more ambitious missions [152,153], which create a need for even more progress.
- These developments will make a lunar outpost more affordable [9] so eventually it will be established, provisioned by lunar mining [153-161]. Space agencies will fund the maturation of key technologies including the manufacture of solar cells [162,163] and structures [30,164-168] from in situ resources.
- Space tourism may establish facilities in space, and they will require provisioning with propellant and spare parts plus other services [169-171].

- If an organization successfully establishes a permanent settlement on Mars, it will become a customer for anything made in space[10] [172]. Because of the Martian gravity, Mars will not become an exporter of manufactured goods for a long time and therefore will not directly contribute to in-space industry, but it will be a consumer of that industry, providing cash flow for the space development companies so they can make additional progress. Martian settlers will also develop mining and manufacturing technologies that have dual-use for the in-space supply chain.
- With these things happening in space, and with the robotics revolution continuing on Earth, it will become clearer to government leaders and the public that space industry is possible, highly desirable, and affordable. Then, government leaders will take on the mantle of the project and complete it for the good of humanity.

*6.2 Building a Cis-Lunar Water Economy*

Many space settlement advocates believe that mining water for propellant is the crucial element in a strategy to start space industry [147-151, 173-176]. Therefore, it is of paramount importance in Stage 1 to ensure water mining becomes profitable. One of the challenges is funding the establishment of a propellant depot in cis-lunar space, possibly in LEO or GEO or at Earth-Moon Lagrange points. A depot will be needed to receive water from mining companies and distribute it to customers such as national space agencies and commercial space tug operators boosting communication satellites. The depot must include a water electrolysis capability, cryocooling and storage, adequate power generation and radiators for these capabilities, plus the usual spacecraft systems [147,177,178], so its cost will be significant. Capital recovery for the depot will be a significant contributor to the cost of propellant made in space. Technologies that make the cis-lunar water business more profitable include new propulsion that has both high thrust and high specific impulse instead of just one or the other. Higher thrust can boost a communication satellite to GEO more quickly, avoiding the loss of profits (upwards of US $100 million) for a satellite owner during the months it takes a solar electric ion thruster to boost the satellite to GEO. The potential to eliminate that loss makes space-mined propellant more competitive. Higher thrust also allows an asteroid mining spacecraft to complete its round-trip mining mission more quickly, so it can acquire the water at less cost. High specific impulse decreases the amount of mined water that must be expended as propellant by the water-mining spacecraft, increasing the net yield, again lowering the cost. Other technologies that will make it more profitable include lower mass systems for propellant depots, faster or more efficient mining methods, and better methods to create space storable propellant like methane or isopropyl alcohol, which can be made with the carbon dioxide that is released by heated carbonaceous asteroid material or that is found in lunar ice.

An important way that space science can contribute to this process is by selecting missions that utilize space resources. For example, I have been working with Honeybee Robotics to develop CubeSats that mine water from an asteroid or the icy surface of a moon like Europa then use the water immediately for steam propulsion [179]. It can use the water to fly around in milligravity or hop on a body like Europa in higher gravity. The reason we proposed this project is because steam propulsion has dual-use; demonstrating it on science missions will help prove to investors the viability of the cis-lunar water economy. Our modeling shows that steam hopping works extremely well [180]. Prototypes of the small mining devices were built by my collaborators and they successfully extracted water from physically dry asteroid regolith simulant using realistic mineralogy for a petrological class 2 or 3 asteroid. The next step is to mature the steam propulsion technology through CubeSat test flights. ISS or sounding rockets can be used for tests of the mining apparatus in zero gravity. Finally, a flight version can be built for a mission to an asteroid or small moon, using mined water to never run out of propellant and to keep exploring. (There are ways to get the necessary delta-v to deliver a CubeSat to these targets.) Already in working this project we have conceived of new technologies for use in cis-lunar space.

---

[10] Mars is too far from Earth to be the locus of bootstrapping a complete supply chain.

A significant fraction of the space development community is working on technologies and business plans related to the cis-lunar water economy. NASA, with a view to buying propellant, is funding businesses and academia to develop the technologies [181-183]; they are also developing such technologies in-house [23,184-186]. NASA's annual Robotic Mining Competition [187] offers incentive points for robots that dig deeper into the regolith, simulating the process of acquiring lunar or Martian water from beneath a desiccated overburden, thereby crowdsourcing the development of digging techniques. At least three CubeSat lunar orbiter missions [188-190] and two lunar lander missions [185,191] are being funded or developed by national space agencies to prospect water on the Moon.

In all this effort, the science community (including the human spaceflight exploration community) is working synergistically with the space development community because their objectives overlap. Space agencies will benefit by having reduced cost for exploration missions so they can do more with the budgets they have. Science benefits because funding to characterize water resources leads to better understanding of solar system evolution. Commercial space companies benefit because the technologies they need are being developed. Ultimately, all humanity will benefit because a cis-lunar water economy makes it less expensive to set up a lunar outpost where robots can begin mining and manufacturing, leading to elements of a supply chain, making it more obvious for policymakers that the vast material resources of the Moon can be used to solve economic and environmental problems on Earth.

7.  **Conclusion**

I have argued that the bootstrapping of a Self-supporting, Replicating Space Industry is technologically feasible and that there is no reason to delay beginning. It will represent a transition in the form of our civilization, freeing us from the constraints of a single planet and giving us great benefits on Earth, both economic for humanity and environmental for all species, as well as making our existence in the universe more secure. It will also revolutionize science and exploration by providing material support and greater access to the space environment. The process of establishing this industry will take several decades. Spreading the costs over that period, humanity can easily fit it into our space science and exploration budgets if we have public support and visionary policymakers who make it a priority. However, the concept is abstract and just expensive enough that it may be difficult to convince them. The space community has been evolving strategies toward space development, and a number of themes are emerging, the most important of which is establishing a cis-lunar water economy by mining asteroids and the Moon. Successfully doing that will lead to more activity in space, both scientific missions led by national space agencies and commercial activity. This will provide the opportunity to develop more technologies for space industry and to demonstrate some limited portions of an in-space supply chain. As these developments occur, and as robotics in terrestrial industry continues its revolutionary trajectory, eventually policymakers should embrace the vision to establish space industry. However, time is of the essence so that space industry may help avert the economic and environmental perils on Earth.

**Acknowledgements**

Some of the research mentioned in this article was supported by the NASA's Small Business Innovation Research and Small Business Technology Transfer (SBIR/STTR) program, Regolith Resource Robotics, contract number NNX15CK13P, "The World is Not Enough (WINE): Harvesting Local Resources for Eternal Exploration of Space."